\documentclass[sigconf]{acmart}
\AtBeginDocument{%
  \providecommand\BibTeX{{%
    \normalfont B\kern-0.5em{\scshape i\kern-0.25em b}\kern-0.8em\TeX}}}

\copyrightyear{2021}
\acmYear{2021} 
\setcopyright{acmcopyright}
\acmConference[UMAP '21] {Proceedings of the 29th ACM Conference on User Modeling, Adaptation and Personalization}{June 21--25, 2021}{Utrecht, Netherlands}
\acmBooktitle{Proceedings of the 29th ACM Conference on User Modeling, Adaptation and Personalization (UMAP '21), June 21--25, 2021, Utrecht, Netherlands}

\acmPrice{xx.xx}
\acmDOI{xx.xxxx/xxxxxxx.xxxxxxx}
\acmISBN{xxx-x-xxxx-xxxx-x/xx/xx}

\usepackage[linesnumbered,lined,ruled]{algorithm2e}
\usepackage{booktabs}
\usepackage{titlesec}
\usepackage{multirow}
\usepackage[normalem]{ulem}
\usepackage{hyperref}
\newcommand{\algoname}[1]{\textsc{{#1}}}
\newcommand{\metricname}[1]{\textit{#1}}
\newcommand{\groupmetricname}[1]{\textit{#1}}
\newcommand{\dave}{\algoname{Dave}}
\newcommand{\chain}{\algoname{The Chain}}
\newcommand{\likeable}{\metricname{likeable}}
\newcommand{\hq}{\metricname{high-quality}}
\newcommand{\understandable}{\metricname{understandable}}
\newcommand{\si}{\metricname{sparked-interest}}
\newcommand{\funny}{\metricname{funny}}
\newcommand{\informative}{\metricname{informative}}
\newcommand{\creative}{\metricname{creative}}
\newcommand{\ww}{\metricname{well-written}}
\newcommand{\valence}{\groupmetricname{valence}}
\newcommand{\content}{\groupmetricname{content}}
\newcommand{\textm}{\groupmetricname{text}}

\begin{document}

\title{Generating Interesting Song-to-Song Segues With Dave}

\author{Giovanni Gabbolini}
\affiliation{
\institution{Insight Centre for Data Analytics \\
School of Computer Science \& IT}
\country{University College Cork, Ireland}
}
\email{giovanni.gabbolini@insight-centre.org}

\author{Derek Bridge}
\affiliation{
\institution{Insight Centre for Data Analytics \\
School of Computer Science \& IT}
\country{University College Cork, Ireland}
}
\email{derek.bridge@insight-centre.org}

\begin{abstract}
    We introduce a novel domain-independent algorithm for generating interesting item-to-item textual connections, or segues. Pivotal to our contribution is the introduction of a scoring function for segues, based on their `interestingness'. We provide an implementation of our algorithm in the music domain. We refer to our implementation as \dave{}. \dave{} is able to generate 1553 different types of segues, that can be broadly categorized as either informative or funny. We evaluate \dave{} by comparing it against a curated source of song-to-song segues, called \chain{}. In the case of informative segues, we find that \dave{} can produce segues of the same quality, if not better, than those to be found in \chain{}. And, we report positive correlation between the values produced by our scoring function and human perceptions of segue quality.
    The results highlight the validity of our method, and open future directions in the application of segues to recommender systems research.
\end{abstract}

\ccsdesc[300]{Human-centered computing~Empirical studies in HCI}
\ccsdesc[500]{Information systems~Recommender systems}

\keywords{segues, user studies, recommender systems, interestingness}

\maketitle

\begin{figure}
 \includegraphics[width=\columnwidth]{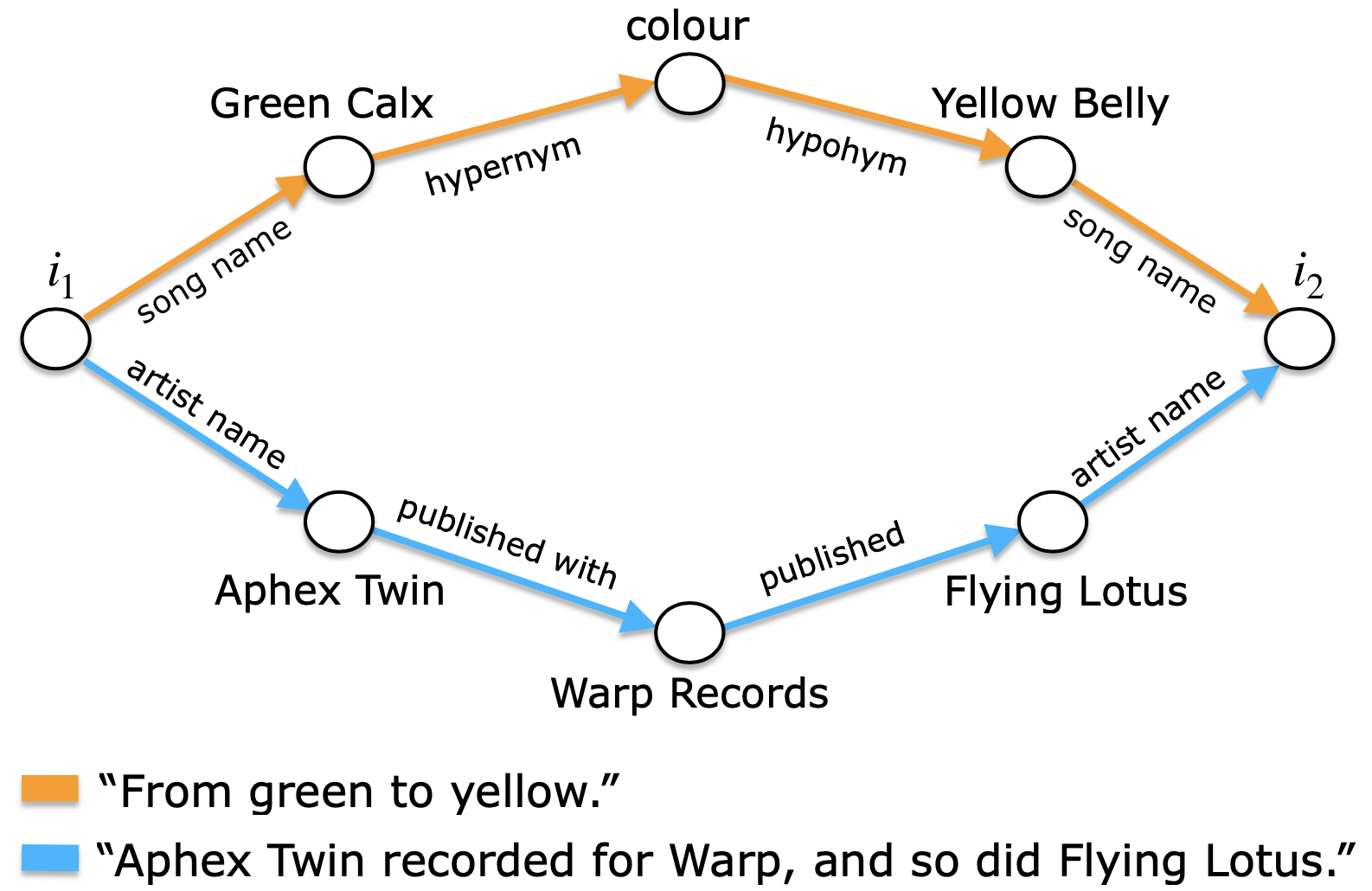}
 \caption{Examples of segues from the song ``Green Clax'' by ``Aphex Twin'' to the song ``Yellow Belly'' by ``Flying Lotus'', represented in the form of paths and texts. 
 }
 \label{fig:teaser}
\end{figure}





\section{Introduction}
Often a recommender system makes use of relationships between items. Knowledge of these relationships and can be used in the recommendation task itself and can enable the explanation of the recommendations. However, there remains much room for research into the discovery of these relationships and how to measure their strength. In \cite{BehroozMTKC19}, Behrooz et al.\ propose the concept of ``segues''. Segues are short texts that explicitly connect one item to another. In \cite{BehroozMTKC19}, segues are used as a mean of enhancing the user experience with voice assistants in music streaming services. The authors provide a simple prototype which is able to generate song-to-song segues, and they evaluate the prototype qualitatively, with the goal only of exploring the potential of the idea.



In this paper we elaborate on the concept of segue, and we provide a domain-independent algorithm for generating interesting item-to-item textual connections. A distinguishing feature of our work is its ability to highlight interesting segues, according to a novel theory of interestingness. 

Our algorithm assumes a knowledge graph as an abstract representation for items and
information about items.
In our abstraction, segues are paths from one item to another, and $\mathit{interestingness}$ is a scoring function for paths. We `get back' from the abstraction by mapping paths to texts.

We implement the algorithm in the music domain. We refer to our implementation as \dave{}. \dave{} can generate song-to-song segues of 1553 different types. Segue types range from factual to word-play.
See Figure \ref{fig:teaser} for examples.

We evaluate \dave{} qualitatively by means of a user trial, where we compare \dave{}'s segues against curated segues from a segment of the Radcliff \& Maconie Show on BBC Radio 6 called \chain{}. In the case of factual segues, we find that \dave{} can produce segues of the same quality, if not better, than those to be found in \chain{}.
The results highlight the validity of our method, and open future directions in the applications of segues to recommender systems research.
We release the source code and the dataset, consisting of the answers gathered during the user trial, to facilitate future research on the subject.\footnote{\url{https://github.com/GiovanniGabbolini/dave}}


The remainder of this paper is organised as follows: Section \ref{sec:related_works} frames segues in the literature on recommender systems and user interfaces for music retrieval, and provides an overview of interestingness measures in data mining. Section \ref{sec:method} explains how \dave{} works. Section \ref{sec:experiments} details the experimental procedure and analyses the results. Section \ref{sec:conclusions} discusses conclusions and future work.

\section{Related Work} \label{sec:related_works}


\subsection{Explanations and recommender systems}
Explanations of recommendations is an active research topic that has helped to increase the value of recommender systems, guided by goals that go beyond recommendation accuracy \cite{Tintarev2011}.
We refer the reader to the paper by Nunes and Jannach \cite{Nunes2017} for a comprehensive survey on the subject.
Vig et al. \cite{Vig2009} divide explanations into three categories: feature-based (``We have recommended this item because of your interest in this feature''), item-based (``We have recommended this item because you like this other item'') and user-based (``People who liked this item also liked this other item''). 
Segues are strongly related to both item-based and feature-based 
explanations. In fact, segues make item-to-item connections explicit by leveraging item features.  They are therefore related to the item- and feature-based hybrid explanations described in \cite{Papadimitriou2012}. 

There is a growing body of research that uses knowledge graphs for computing recommendations \cite{Musto20191, Musto2018, Musto2017, Bellini2017, Ostuni2014, DiNoia2012, DiNoia2016} 
and some research that then uses the knowledge graph for  explanations of those recommendations 
\cite{Passant2010, Musto2019, Musto2016, Bellini2018, Zhao2020,Koettgen2020}.
For example, in \cite{Musto2019, Musto2016} the authors explain recommendations by jointly representing the following in a knowledge graph: items liked by a user; items recommended to the user; and features mined from DBPedia. The candidate explanations are paths from the user profile to the recommended items, which are ranked and translated into natural language.
Bellini et al.\ \cite{Bellini2018} propose an autoencoder neural network for explainable recommendation. The hidden layer, and its connections to the input and output, are replaced by a knowledge graph. The graph is mined from DBPedia, and contains item features. In parallel with recommendations, the model also delivers explanations by leveraging the weights learned for the graph edges.

In some recommenders, the items to be recommended are inferred from paths, and the same paths are also used for explanations.
Passant \cite{Passant2010}, for example, proposes a heuristic for scoring paths by relevance, and recommends items associated with the most relevant paths. In \cite{Zhao2020,Koettgen2020}, paths are scored through the use of deep neural networks that learn from past user interactions. 
We consider our work to be closely related to path-based recommenders: we introduce a scoring function for paths 
that can be used for recommendation and explanation purposes. Our scoring function is novel because it strives to maximize the quality of segues. 


\subsection{User interfaces for music retrieval}
The study of intelligent user interfaces for music retrieval has been an active research topic for twenty years now \cite{Knees2020}.  The main goal of early interfaces was to facilitate the browsing of personal music collections, for example, by representing songs in 2D or 3D maps. Gulik and Vignoli \cite{Gulik05} visualize artists as dots in a 2D map where proximity is determined by similarity across attributes such as genre, tempo and year; Tzanetakis et al. \cite{Tzanetakis2001} visualize songs as dots in a 3D map, arranged by running PCA on song features, and where the genre of a song determines the colour of its dot.
Another approach consists in representing collections linearly. Pohle et al. \cite{Pohle2007}, for example, arrange songs circularly by solving a Travelling Salesman Problem, where the distances are determined by song-to-song similarity.

Recently, with the advent of streaming services, the quantity of music available to listeners shifted from relatively small personal collections to, virtually, all music content. Modern interfaces are, for the most part, not built to handle the visualization of these very large collections, but to automatically search for personalized sets of songs, and to show just these to the user \cite{Knees2020}. 
In Pampalk et al. \cite{Pampalk2007}, for example, new artists are recommended based on a seed artist picked by the user. The recommendation algorithm is based on artist-to-artist similarity. The seed song is visualized as a sun, and the recommendations as sun rays, labelled by keywords.

The idea of segues is complementary to the work on user interfaces for music retrieval as proposed in the literature. These interfaces use similarity to facilitate browsing, search and recommendation \cite{Knees2016}. Segues, on the other hand, make the connections explicit.
Segues were first introduced in \cite{BehroozMTKC19}. There, the authors developed a simple prototype able to generate segues for consecutive songs in playlists. They score individual segues simply with static segue preference weights but they also take into account, for example, the position of the segue in the playlist and its proximity to segues of the same type, in order to obtain a degree of segue diversity. Their experimental procedure consisted of unstructured interviews, aimed at exploring the potential of the idea. Our contribution fills a number of gaps in this previous work. We develop interestingness, a scoring mechanism for individual segues. And, we evaluate our system with a larger user trial, comparing it with a curated baseline.

\subsection{Interestingness measures for data mining}
Data mining is the discovery of patterns in large and complex datasets \cite{Hand2015}. 
Measuring the interestingness of discovered patterns is an active and important field of study, surveyed by Geng and Hamilton \cite{Geng2006}. Most research in this direction has focused on the interestingness of certain kinds of patterns, such as association rules, classification rules and summaries. Geng and Hamilton highlight that interestingness is usually defined as a combination of different components, such as: conciseness (rewarding patterns that contain a relatively small number of elements); peculiarity or infrequency (rewarding patterns if they are far away from other discovered patterns, according to some distance measure); surprisingness (rewarding patterns that contradict a person's existing knowledge or expectations); and actionability in some domain (rewarding patterns if they enable decision-making about future actions in this domain).

There is a smaller amount of research on the interestingness of paths in graphs. Lin and Chalupsky \cite{Lin2003}, for example, propose that the interestingness of a path of a given type is determined by the infrequency of that type of path in the graph. 
Ramakrishnan et al. \cite{Ramakrishnan2005} leverage three heuristics for interestingness. The first implements the idea that more specific nodes and edges convey more information than general ones, e.g.\ ``singer'' is more specific than ``person''. They compute specificity from hierarchies of node and edge types. Their second heuristic favours paths that cross different domains, e.g.\ from music to cinema. For this, a manual labelling, assigning domains to the nodes in the graph, is required. The third takes into account the infrequency of the path type.
Aleman-Meza et al.\ \cite{Aleman-Meza2003} also introduce three heuristics for interestingness. The first implements the idea of conciseness, and is computed from path length. The second is a heuristic for user-personalized paths. It leverages personalized weights assigned to regions of the graph. The third is concerned with trust in data sources. Again in this case, trust weights are manually assigned to regions of the graph.

\section{Method} \label{sec:method}
Our goal is to generate interesting item-to-item textual connections, or ``segues''. That is, we aim for interestingness, and we discourage trivial and boring segues. We are interested in generating factual connections, but also in amusing ones, based on simple word-play. 

\subsection{Algorithm} \label{subsect:algorithm}
Our algorithm uses a knowledge graph $G$ as an abstract representation for items and information about those items.
In our abstraction, segues are paths from an item to another item, and interestingness is a scoring function from paths to numbers. We `get back' from the abstraction through another function, which maps paths to texts.
In particular, the algorithm for finding a segue from an item $i_1$ to another item $i_2$ works as follows: we find 
the paths in $G$ that connect $i_1$ to $i_2$, we score them based on their interestingness, keeping only the best one, which is translated to text, and finally returned. This approach allows us to exploit heterogeneous data \cite{Fensel2020}, and is domain-independent.
The algorithm is summarized as Algorithm \ref{alg:1}. 
\IncMargin{1em}
\begin{algorithm}
\SetKwInOut{Input}{input}\SetKwInOut{Output}{output}
\Input{\,knowledge graph $G$, items $i_1$ and $i_2$}
\Output{\,interesting segue}
\BlankLine
paths = $\mathit{find\_paths}(G, i_1, i_2)$ \\
\For{\textnormal{path in paths}}{
    path.score = $\mathit{interestingness}$(path)
}
best$\_$path = path with highest score \\
best$\_$segue = $\mathit{path\_to\_text}$(path) \\
\textbf{return} best$\_$segue
\caption{$find\_segue(G, i_1, i_2)$}
\label{alg:1}
\end{algorithm}
\DecMargin{1em}

In the following, we will provide details for Algorithm \ref{alg:1}. We start by presenting  some preliminary concepts:
\begin{description}
\item[\textnormal{A} knowledge graph] is a set of triples $G=\{(e,r,e')\,|\,e,e' \in E, r \in R\}$, where $E$ and $R$ denote, respectively, the sets of entities and relationships. A special subset of entities $I \subseteq E$, are the items (in our case, songs). Every entity has a \textit{type} and a \textit{value}. 
For example, an entity that represents a song has \textit{type} equal to ``song'' and a \textit{value} equal to the song URI. Every relationship has a \textit{type}.
\item[\textnormal{A} path] $p$ in $G$ is an ordered list of entities and relationships in $G$, $p=[e_1, r_1, ..., r_{n-1}, e_n]$ where each triple in $p$ must be in $G$. The \textit{type} of $p$ is
the ordered concatenation of the entity and relationship types in $p$. 
\end{description}
We turn to the components of Algorithm \ref{alg:1}, namely $\mathit{find\_paths}$,  $\mathit{path\_to\_text}$ and $\mathit{interestingness}$.
$\mathit{find\_paths}$ is a 
path-finding procedure that returns all the simple paths, i.e.\ those without cycles, starting from the item $i_1$ and reaching the item $i_2$ in $G$ . However, we allow for the possibility of constraining the paths that $\mathit{find\_paths}$ can find; see Section \ref{sec:alg_mod}. 
$\mathit{path\_to\_text}$ works by template filling, with canned templates indexed by the path type, and completed with information on the entities and relationships that constitute it. Our focus for this work was mainly on $\mathit{interestingness}$ and so we devote the remainder of this section to explaining our definition.


Defining a scoring function for segues based on interestingness is not a trivial task. To be concrete, we refer to Figure \ref{fig:teaser}: which one of the two segues is more interesting? There is no correct answer to this question, as interestingness can depend upon personal relatedness \cite{Schank1979} and background knowledge \cite{Kintsch1980}.
In this context, we develop a simple theory of interestingness according to which a ranking can be determined. Our theory builds upon the concepts of infrequency and conciseness. We believe that infrequent segues are more interesting, as pointed out by Schank \cite{Schank1979} and Kintsch \cite{Kintsch1980} when discussing interestingess of general statements. We also believe that concise segues are more interesting, as supported by Geng et al.\ \cite{Geng2006} in the context of interestingness in data mining. 
Our definition of interestingness applies to paths in knowledge graphs and consists of three heuristics. 
The heuristics rely only on statistical information and simple content descriptors, that can be defined independently of the domain, as they do not depend on the semantics of segues.

\begin{description}
 \item[Rarity] 
 We define the rarity of a path $p$ using the proportion of paths in $G$ that have the same type as $p$. To formalize this, let $T$ be the set of all path types in $G$; and let $f(t)$ be the number of paths in $G$ that are  of type $t$. Then,
\[\mathit{rarity}(p)=1-\frac{f(\mathit{type}(p))}{\max_{t \in T} f(t)}\] 

\item[Unpopularity]
We define the unpopularity of a path $p$ using the notion of centrality of an entity $e$. An entity is central if it has a high number of incoming and outgoing edges, compared with other entities of the same type. A path that visits central entities is more popular than one that does not. To formalize this, let $\mathit{edgeset}(e)$ be the set of incoming and outgoing edges to and from an entity $e \in E$ in $G$.
We define the centrality of an entity $e$ as: 
\[\mathit{centrality}(e)=\min \left( 1, \frac{|\mathit{edgeset}(e)|}{\underset{e' \in E}{\mathit{median}} |\mathit{edgeset}(e')|} \right)\, , \, \mathit{type}(e')=\mathit{type}(e) \]
Then, we define the $\mathit{unpopularity}$ of a path $p$ as:
\[
    \mathit{unpopularity}(p)=1-\min_{e \in p \cap E}(\mathit{centrality}(e))
\]

\item[Shortness] Let the $\mathit{length}$ of a path $p$ in $G$ be:
\[\mathit{length}(p)=|p \cap R|\]
We define the $shortness$ of a path $p$ in $G$ as done in \cite{Aleman-Meza2005}:
\[
    \mathit{shortness}(p)=\frac{1}{\mathit{length}(p)}
\]
\end{description}
The heuristics $\mathit{rarity}$ and $\mathit{unpopularity}$ both implement the idea of favouring infrequent segues, but in a different fashion: $\mathit{rarity}$ has high values for infrequent path types, while $\mathit{unpopularity}$ has high values for paths that include infrequent entities.
For example, a path that connects people who share a birthday is likely to have a fairly high value for $\mathit{rarity}$, but not necessarily for $\mathit{unpopularity}$, e.g.\ if both people are very famous.
Or, paths that connect people who have the same hair colour will have a low value for $\mathit{rarity}$, but can have a high value for $\mathit{unpopularity}$, e.g.\ if the shared hair colour is something unusual such as ``green''.
The $\mathit{shortness}$ heuristic implements the concept of conciseness.

We define the $\mathit{interestingness}$ of a path $p$ in $G$ as the convex combination of the three heuristics:
\[
    \mathit{interestingness}(p)=w_1 \mathit{rarity}(p) + w_2 \mathit{unpopularity}(p) + w_3 \mathit{shortness}(p)
\]
It ranges from zero to one.
$w_1, w_2, w_3$ are parameters to be tuned, subject to $w_1+w_2+w_3=1.$

\subsection{Implementation}
We implement Algorithm \ref{alg:1} as described in Section \ref{subsect:algorithm} for the music domain. We consider items to be songs, and obtain song-to-song segues as a result. We will refer to our implementation from now on as \dave{}.

\subsubsection{Knowledge graph}
We represent a song with three fields: \textit{song name}, \textit{artist name} and \textit{album name}. The representation allows missing values, e.g.\ a song that is not part of an album, and can be easily changed with only minor modifications to the rest of our implementation, e.g.\ if songs were instead represented by their Spotify URIs. 

Our implementation uses a knowledge graph with 40 distinct node types and 230 distinct edge types, and can provide segues of 1153 different path types.
We build the knowledge graph with data that we harvest from three main resources:

\begin{description}
 \item[MusicBrainz] We use MusicBrainz\footnote{\url{https://musicbrainz.org/}} as the main source of factual data. We exploit the MusicBrainz APIs.\footnote{\url{https://python-musicbrainzngs.readthedocs.io/en/v0.7.1/}} They allow us to navigate the MusicBrainz database, and offer entity-linking functionalities. In a first step, we link the actual song, its album and its artist to their respective MusicBrainz URIs. Then, we mine different sorts of factual data, ranging from the genres of the song to the birth places of the artists.

\item[Wikidata] We use Wikidata\footnote{\url{https://wikidata.org/}} as an additional source of factual data. There exists a mapping from MusicBrainz URIs to Wikidata URIs, making it easy to use both resources. From Wikidata, we mine biographical data about artists that is not available in MusicBrainz, e.g.\ the awards that an artist has won.

\item[WordNet] We use WordNet \cite{Fellbaum2000} to mine lexical data about the words in song, artist and album names, e.g.\ hypernyms and hyponyms.
\end{description}
In addition, we gather some further lexical data through a number of different resources. For example, we link words in song, artist and album names to their stems using the Porter stemmer \cite{Porter1980}, and to their phonetics with the NRL algorithm \cite{Elovitz1976}. We provide a complete description of entities and relationships that build up the knowledge graph in the additional materials.\footnote{\url{https://doi.org/10.5281/zenodo.4619395}} 

The factual data that we obtain from MusicBrainz and Wikidata allows for conventional, informative segues. On the other hand, the lexical data from WordNet and other resources yields less conventional and perhaps amusing connections based on word-play. We show two examples in Figure \ref{fig:teaser}.

\subsubsection{Algorithm} \label{sec:alg_mod}
As mentioned in Section \ref{subsect:algorithm}, we constrain the $\mathit{find\_paths}$ component of Algorithm \ref{alg:1}; specifically, we constrain it to find only paths that go from a start song to another song without visiting another song. This constraint limits the number of paths to be scored, at the price of losing some indirect paths. We believe that some of these indirect paths would be filtered out by $\mathit{interestingness}$ anyway, since they would have low values for $shortness$.
We set the weights to be used in the $\mathit{interestingness}$ score to $w_1=0.4, w_2=0.2, w_3=0.4$. The weights were set after empirical experimentation, using songs that were not used in the user trial. We discuss other ways to set the weights in Section \ref{sec:conclusions}. 

\section{Experiments} \label{sec:experiments}
We evaluate \dave{} by means of a user trial, where we compare \dave{}'s segues against a curated source of segues called \chain{}. \chain{} is a segment of the Radcliffe \& Maconie Show on BBC Radio 6. In this segment of the show, listeners call in and propose the next song, always making sure that there is a connection (sometimes informative, sometimes funny) between the previous song and the next one. \chain{} is made for entertainment and therefore offers very interesting segues, with strong creative traits. A database with all the segues that have appeared in \chain{} is available on the internet.\footnote{\url{https://www.thechain.uk}} At the time of writing, it comprises more than $8000$ segues.
A direct comparison with curated segues is our means of assessing the quality of segues from \dave{}. We do not include any other algorithmic baseline, as we are not aware of any other similar systems in the literature. The only work that deals with segues is \cite{BehroozMTKC19}. 
We cannot compare with their work 
their scoring mechanism is presented in insufficient detail to be reproducible. 

Our user trial is a within-subject trial with two treatments: \dave{} and \chain{}.
We generate segues to show in each treatment using the following procedure. One segue is sampled at random from a fixed random sub-sample of \chain{}, of cardinality equal to 200 segues. The corresponding segue is generated by \dave{} by picking the segue that maximizes the $\mathit{interestingness}$, when starting from the same song as \chain{}, and going towards a song from a fixed sample of songs, of cardinality equal to 496 songs. These 496 songs are a random sub-sample of 20000 songs from the RecSys Challenge 2018 Dataset \cite{Chen2018}. We use MusicBrainz as a source of artist data to filter out artists based on the country in which they were born or in which they are based, and on the musical genres with which they are associated. Specifically, we keep only artists born or based in the UK, and who are associated with at least one genre from the following: blues, blues-rock, pop, pop-rock, rock, soft-rock, funk, jazz, r\&b and soul. This country and these genres are ones that predominate in \chain{}. This filtering ensure that the songs that \dave{} can choose from match the style of songs found in \chain{}, thus mitigating one confounder from the user trial.

Each participant is asked to evaluate six segues: she undergoes both treatments  three times. The order of treatments in each pair is randomized. We make sure that the same segues are not shown multiple times to a participant.

In the following, we provide details on the user trial design, we present statistics on the answers, and, finally, we analyze the results in depth.

\subsection{User trial design}
The user trial begins with an instructions page. It continues with a three-part survey, whose parts we will refer to as intro, segue evaluation and outro. The trial concludes with a final page that offers an optional comments box. In the following, we describe the three parts of the survey. We provide screenshots of the user trial text and its workflow in the additional materials.\footnote{\url{https://doi.org/10.5281/zenodo.4619395}}


\subsubsection{Intro}
In the intro, we ask each participant some questions to identify how much she engages with music. A previous study has highlighted that segues might be especially suited for music `nerds' \cite{BehroozMTKC19}. By asking these questions in our trial, we can see whether music engagement correlates with segue appreciation, for example. The source of the questions that we ask is the Goldsmiths Musical Sophistication Index (Gold-MSI) \cite{Mullensiefen2014}. The index comprises five aspects. Four of the aspects are concerned with music skills, such as musical training and singing abilities. We restrict ourselves to the remaining aspect, the one called Active Engagement (AE). AE covers ``a range of active
musical engagement behaviours (e.g.\ I often read or search the internet for things related to music) as well as the deliberate allocation of time and money on musical activities (e.g.\ I listen attentively to music for $n$ hours per day)'' \cite{Mullensiefen2014}. We follow \cite{Mullensiefen2014}, and we measure AE by asking ten questions, with answers on a seven point scale.

We are also interested in English proficiency, as we want to make sure that participants can properly understand the segues. So in the intro we also ask the participant to identify her level of proficiency from among four options: ``Low'', ``Mid'', ``High'' and ``Mother Tongue''. 
We do not collect demographic information, as we considered it not essential for the scope of the study.

\subsubsection{Segue evaluation}
In the segue evaluation phase, each participant is asked to evaluate six segues, as described in Section \ref{sec:experiments}. 
We show the segues, as well as the titles and artists of the songs that they connect. We do not provide any means for the user to listen to the songs. For every segue, we ask questions to measure a variety of quality metrics. First, we ask whether the segue is \likeable{}, of \hq{}, and whether it \si{} in the next song. Second, we asked how the participant perceived its content, on three dimensions: \informative{}, \funny{} and \creative{}.
Lastly, we wanted to make sure that the connection between the two songs is expressed in an \understandable{} way by the segue, and that the segue is \ww{}.
We call these three groups of dependent variables respectively: \valence{}, \content{} and \textm{} quality metrics. The dependent variables introduced
in this part of the survey are summarized in Table \ref{table:dep_variable}. We measure all of them using five point Likert scales. 

\begin{table}
\caption{Dependent variables measured during the segue evaluation part of the user trial.}
\begin{tabular}{ll}
\toprule
name & dependent variables \\
\midrule
\valence{} quality metrics             &     \likeable{}, \hq{}, \si{} \\
\content{} quality metrics           &     \funny{}, \informative{}, \creative{} \\
\textm{} quality metrics             &     \understandable{}, \ww{} \\
\bottomrule
\end{tabular}
\label{table:dep_variable}
\end{table}

\subsubsection{Outro}
In the outro part of the trial, we measure the familiarity of the participant with the songs and artists involved in the segues that she has evaluated. Familiarity with songs and artists might be a confounder for the quality metrics, and we want to address these effects, if any. We consider familiarity because it has been shown to be an accurate predictor of musical choice, at least as good as liking \cite{Ward2014}, and we assume there may be an extension of the results in \cite{Ward2014} to segues.  We measure familiarity with the songs and artists of both the first song and second song in each segue. Familiarity is measured with a simple two point scale: ``Familiar'' and ``Not familiar''. We decided to ask about familiarity in the outro, after the segue evaluation, so that we avoid accentuating any confounding effects.

\subsection{Answer statistics} \label{subsect:data-preprocessing}
In total, 158 people completed the trial. They are undergraduate Computer Science students recruited in a university in Ireland.
The median completion time for the survey is 7 minutes, with a maximum of 68 minutes and minimum of 2 minutes. We discard answers from people who took less than 3 minutes, as we are worried about their reliability. No participant declared their English proficiency to be ``low'' but we further filter out participants with ``mid'' English proficiency, since they also might not be able to properly evaluate segues. We are left with people with ``high'' proficiency and those for whom English is their mother tongue (90\% of the total). After the filtering, we end up with 151 people. We convert Active Engagement answers to numbers from one to seven, and segue evaluation Likert scale answers to numbers from one to five.
We also analyze the comments left by the participants, 35 comments in total, but we do not discuss them in this paper for lack of space and because they do not add much to our analysis.

\paragraph{User categories}
In some of our analysis, we partition participants based on their level of Active Engagement with music (AE). We summed the answers to the AE questions given by each participant, obtaining a distribution of total AE scores. We divided participants according to the quartiles of this distribution, giving four categories: ``low-AE'', ``mid-low-AE'', ``mid-high-AE'' and ``high-AE''. We validated this partition by considering confidence intervals of segue evaluation answers: the quartiles show good internal cohesion.

\paragraph{Segue categories} 
We believe that segues can be divided into two categories: those that are intended to amuse (``funny'') and those that are intended to impart information (``informative''). 
We decided to create a ground truth that assigns each segue to one of the two categories. We, the two authors of this paper, separately labeled every segue manually, guided by the following criterion: a segue is funny if it is written with the goal of making the listener smile, and a segue is informative if it is written with the goal of giving information to the listener. A segue can have both goals, e.g. if it presents information in a funny way. In such borderline cases, we assigned the goal that seemed dominant. We disagreed upon the labelling in 13 out of 400 cases (Cohen's $k=0.92$). We solved divergences as follows: given a segue $s_1$ where there was disagreement, we found a second segue $s_2$ whose category was not in dispute, and that both of us considered to be similar to $s_1$. 
We then assigned to $s_1$ the category of $s_2$. 
We validate the ground truth by double-checking it against the answers to the survey. In particular, participants were asked to express how much segues were \informative{} and \funny{}. We computed mean values of their answers, considering separately segues labelled as informative and funny in the ground truth. We carry out a \textit{t}-test for assessing the significance of differences in the mean values. We found that the mean for \informative{} is statistically significantly higher than the mean for \funny{} for segues labelled as informative ($3.64$ vs $2.81$, $p$$<$$0.001$), and the opposite for segues labelled as funny ($2.47$ vs $2.82$, $p$$<$$0.001$). We conclude that participants in the survey agree with our manual labeling, thus providing some support for the reliability of the ground truth.

The ground truth reveals that \chain{} is biased towards funny segues: roughly three out of every four of its segues are funny. \dave{} is approximately balanced. We show some examples of informative and funny segues in Tables \ref{table:example_segues_informative} and \ref{table:example_segues_funny}.

\setlength{\tabcolsep}{4pt}
\begin{table*}
\caption{Examples of informative segues. Not only were these labeled informative in the ground truth but also at least two user trial participants gave them a maximum rating on the \informative{} quality metric.}
\begin{tabular}{ccccc}
\toprule
treatment  & first song & segue & second song \\ \midrule
\dave{}  & \textit{Weather To Fly} by \textit{Elbow} & \begin{tabular}{c} And now Guy Garvey, \\ who was a member of Elbow... \end{tabular} &  \begin{tabular}{c} \textit{Belly Of The Whale} \\ by \textit{Guy Garvey} \end{tabular}  \\ \midrule
\chain{} & \begin{tabular}{c}
     \textit{One for the Road}  \\
     by \textit{Arctic Monkeys} 
\end{tabular}  & \begin{tabular}{c}
     On Arctic Monkeys’ last tour, Bill \\ Ryder-Jones
 of The Coral joined them...
\end{tabular} & \textit{Pass It On} by \textit{The Coral} \\
\bottomrule
\end{tabular}
\label{table:example_segues_informative}
\end{table*}
\setlength{\tabcolsep}{5pt}

\begin{table*}
\caption{Examples of funny segues. Not only were these labeled funny in the ground truth but also at least two user trial participants gave them a maximum rating on the \funny{} quality metric.}
\begin{tabular}{ccccc}
\toprule
treatment  & first song & segue & second song \\ \midrule
\dave{}  & \textit{Fleety Foot} by \textit{Black Uhuru} & From foot to faces\ldots & \textit{Faces} by \textit{Ed Sheeran} \\ \midrule
\chain{} & \begin{tabular}{c} \textit{Tumbling Dice} by \\ \textit{The Rolling Stones} \end{tabular} & \begin{tabular}{c} You need a dice to play \\ snakes and ladders... \end{tabular} & \begin{tabular}{c} \textit{Rattlesnakes} by \\ \textit{Lloyd Cole and the Commotions} \end{tabular} \\
\bottomrule
\end{tabular}
\label{table:example_segues_funny}
\end{table*}
\setlength{\tabcolsep}{5pt}

\subsection{Results} \label{sec:results}
In this section, we analyze the answers to the segue evaluation part of the survey, dividing by treatments, segue category and user category. We also investigate the effect of familiarity. And, finally, we evaluate the effectiveness of $\mathit{interestingness}$ and the correlation between quality metrics.

\subsubsection{Performance in the quality metrics}
We compute the average for each quality metric given in Table \ref{table:dep_variable} within treatment (\dave{} and \chain{}), presenting separately the results for informative segues (Table \ref{table:quality_informative}) and funny segues (Table \ref{table:quality_funny}). We conduct a \textit{t}-test for assessing the significance of differences between the two treatments. 
We discuss these results below.

\paragraph{Informative segues}
For informative segues (Table \ref{table:quality_informative}), \dave{} outperforms the human-curated segues of \chain{} for two of the three \valence{} quality metrics but the differences are not statistically significant. There are statistically significant differences on the \content{} quality metrics: \chain{} is perceived as more \funny{} and \creative{} ($p$$<$$0.05$), while \dave{} is more \informative{} ($p$$<$$0.01$). Finally, turning to the \textm{} quality metrics, \dave{}'s segues turn out to be better written and more \understandable{} than \chain{}'s but again without statistical significance.

\paragraph{Funny segues}
When it come to funny segues (Table \ref{table:quality_funny}), human-curated segues from \chain{} outperform \dave{}'s segues across all the quality metrics, with statistically significant differences.
We notice low values for \funny{} in both treatments. We might expect it to be higher in the category of segues we are considering. 
This may be due to the medium of presentation of the segues, i.e.\ read on a screen. We might expect better results if, for example, segues were spoken. It may also just be that the word-play humour of these segues does not appeal to the sense-of-humour of the trial participants.


\setlength{\tabcolsep}{4pt}
\begin{table*} 
\caption{Informative segues. Average value of quality metrics achieved in the two treatments, on a scale from one to five. $^*$: $p$$<$$0.05$; $^{**}$: $p$$<$$0.01$.}
 \label{table:quality_informative} 
\begin{tabular}{lcccccccc}
\toprule
{} & \likeable{} &     \hq{} &           \si{} &      \funny{} & \informative{} &    \creative{} & \understandable{} &          \ww{} \\
\midrule
\dave{}  &             $3.26$ &       $3.22$ &       $3.06$ &          $2.39$ &           $3.73^{**}$ &             $3.33$ &                   $3.84$ &       $3.46$ \\
\chain{} &             $3.20$ &       $3.20$ &       $3.13$ &      $2.64^{*}$ &                $3.43$ &         $3.59^{*}$ &                   $3.69$ &       $3.33$ \\
\bottomrule
\end{tabular}
 \end{table*}
 \setlength{\tabcolsep}{5pt}
 
 \setlength{\tabcolsep}{4pt}
 \begin{table*} 
 \caption{Funny segues. Average value of quality metrics achieved in the two treatments, on a scale from one to five. $^*$: $p$$<$$0.05$; $^{**}$: $p$$<$$0.01$; $^{***}$: $p$$<$$0.001$.}
\begin{tabular}{lcccccccc}
\toprule
{} & \likeable{} &     \hq{} &           \si{} &      \funny{} & \informative{} &    \creative{} & \understandable{} &          \ww{} \\
\midrule
\dave{}  &             $2.94$ &        $2.77$ &       $2.76$ &          $2.71$ &                $2.66$ &             $3.22$ &                   $3.58$ &       $3.13$ \\
\chain{} &       $3.28^{***}$ &  $3.14^{***}$ &   $2.99^{*}$ &      $2.89^{*}$ &           $2.90^{**}$ &       $3.57^{***}$ &               $3.78^{*}$ &   $3.35^{*}$ \\
\bottomrule
\end{tabular}
 \label{table:quality_funny} 
 \end{table*}
 \setlength{\tabcolsep}{5pt}

\subsubsection{Correlation of quality metrics}
\begin{figure}
 \includegraphics[width=\columnwidth]{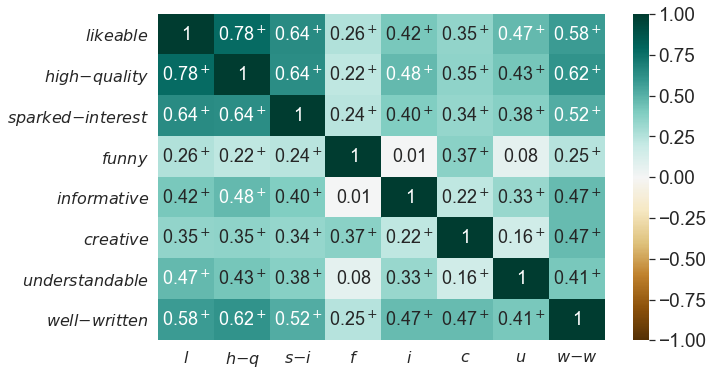}
 \caption{All segues. Quality metrics correlation matrix. $^+$: $p$$<$$0.001$, otherwise $p$$>$$0.05$}
 \label{fig:heatmap}
\end{figure}

We compute pairwise correlations of the eight quality metrics and this is shown for all segues in Figure \ref{fig:heatmap}.
There is high correlation ($0.64$, $p$$<$$ 0.001$) between both \hq{} and \likeable{} with \si{}: good segues can spark interest in the next song.
We notice that \informative{} has higher correlation than \funny{} with all the \valence{} quality metrics: 
a segue perceived as very informative is likely to be also perceived as very likeable, high-quality, and is more likely to spark interest in the next song. This happens to a lesser extent for segues perceived as very funny. Therefore, from a recommender systems point-of-view, where sparking interest is important, it might be more fruitful to address efforts into generating informative segues, as opposed to funny segues.
However, this observation might be just due to the medium of presentation of the segues i.e.\ read. We do not know whether the result would generalize to other mediums, e.g.\ spoken.

The same happens with \understandable{}, which is statistically significantly correlated with \informative{} but not with \funny{}. Further, \creative{} has good correlation with all the \valence{} quality metrics: creativity is a good asset for segues. 
Finally we report high correlation of \ww{} with all \valence{} quality metrics (ranging from 0.52 to 0.62, $p$$<$$ 0.001$). This is expected, since segues are consumed in textual form. How they are written is very important, as important as the content itself.

We have also looked at these correlations in various subdivisions of the data: within treatment (\dave{} or \chain{}), within segue category (funny or informative) and within user category (low-AE, etc.). The narrative that would accompany each of these correlation matrices is not
appreciably different from the one we have given above. Hence, to save space, we do not show these more specific correlation matrices in this paper. 

\subsubsection{Performance in quality metrics and user category}
We divide users into four groups, as explained in Section \ref{subsect:data-preprocessing}, and we compute the average of the answers to the quality metric questions within each user category. We conduct a statistical test for assessing the significance of differences in performance, comparing the lowest level of AE with the other three. We show the results in Table \ref{table:AE}. Only one metric changes statistically significantly: \si{}. This is reasonable: higher active engagement with music is somehow correlated with a propensity for music discovery. The other main quality metrics slightly increase with AE, but the differences are not statistically significant. The results we obtain if we further divide, e.g.\ by treatment or by segue category, confirm those that we have presented in Table \ref{table:AE}. The results contradict the intuition that segues are especially suited to "nerds" \cite{BehroozMTKC19}. The result might change if segues were to include more musicological detail, for example, the synthesizer brand used by two artists during a recording. At present, \dave{}'s knowledge graph does not contains these kinds of details and so does not allow \dave{} to produce segues such as these.

\setlength{\tabcolsep}{3pt}
\begin{table*} 
\caption{All segues. Average value of quality metrics, divided by level of Active Engagement (AE) with music. Values range from one to five. We conduct a significance test, comparing the lowest value of AE with the other three.  $^*$: $p$$<$$0.05$; $^{**}$: $p$$<$$0.01$; $^{***}$: $p$$<$$0.001$.} 
\begin{tabular}{lcccccccc}
\toprule
{} & \likeable{} &     \hq{} &           \si{} &      \funny{} & \informative{} &    \creative{} & \understandable{} &          \ww{} \\
\midrule
low-AE      &             $3.08$ &       $3.00$ &        $2.64$ &          $2.56$ &                $3.07$ &             $3.28$ &                   $3.63$ &       $3.18$ \\
mid-low-AE  &             $3.20$ &       $3.08$ &   $2.97^{**}$ &          $2.72$ &                $3.01$ &             $3.34$ &                   $3.71$ &       $3.23$ \\
mid-high-AE &             $3.19$ &       $3.09$ &  $3.16^{***}$ &      $2.80^{*}$ &                $3.26$ &        $3.54^{**}$ &                   $3.72$ &  $3.46^{**}$ \\
high-AE     &             $3.25$ &       $3.16$ &  $3.09^{***}$ &          $2.66$ &                $3.19$ &        $3.53^{**}$ &               $3.86^{*}$ &   $3.42^{*}$ \\
\bottomrule
\end{tabular}
 \label{table:AE} 
 \end{table*}
 \setlength{\tabcolsep}{5pt}

\subsubsection{Performance in quality metrics and familiarity}
We consider whether quality metrics are related or not to familiarity with the artists and songs involved in the segues. 
For lack of space, we focus our attention on familiarity with songs, stating only briefly the results we have for artists. We divide answers into four groups, based on whether participants are familiar or not with each of the two songs connected by the segue, and we compute the averages of each group. We conduct a statistical test for assessing the significance of differences in performance, comparing the first group (familiar with neither song) with the other three. We do not further partition by treatment, segue category or user category, as the cardinality of some of the groups is already small. We show the results in Table \ref{table:familiarity}.

We observe that familiarity with songs, in general, leads to higher appreciation of segues. Segues are more \likeable{} when connecting two familiar songs than when connecting two unfamiliar songs ($p$$<$$0.01$). And, segues are able to spark interest more when the songs are already familiar, with respect to when they are not ($p$$<$$0.001$). Moreover, they are perceived as better written ($p$$<$$0.01$), and more \understandable{} ($p$$<$$0.05$).

When repeating the analysis but considering familiarity with the artists, we observe the same phenomena, but the increases in the metrics have lower magnitudes. We conclude that familiarity with artists is a weaker confounder than familiarity with songs.

\setlength{\tabcolsep}{3pt}
\begin{table*} 
\caption{All segues. Average value of quality metrics, dividing answers based on the familiarity with the two songs connected by the segue. Values range from one to five. We conduct a significance test, comparing the first group (familiar with neither song) with the other three.  $^*$: $p$$<$$0.05$; $^{**}$: $p$$<$$0.01$; $^{***}$: $p$$<$$0.001$.} 
\begin{tabular}{lcccccccc}
\toprule
{} & \likeable{} &     \hq{} &           \si{} &      \funny{} & \informative{} &    \creative{} & \understandable{} &          \ww{} \\
familiar with      &                    &              &               &                 &                       &                    &                          &              \\
\midrule
neither song      &             $3.11$ &       $3.03$ &        $2.81$ &          $2.65$ &                $3.08$ &             $3.36$ &                   $3.67$ &       $3.25$ \\
just $1^{st}$ song &             $3.22$ &       $3.08$ &    $3.08^{*}$ &          $2.64$ &            $3.30^{*}$ &             $3.45$ &                   $3.87$ &       $3.40$ \\
just $2^{nd}$ song &             $3.29$ &       $3.18$ &  $3.27^{***}$ &          $2.79$ &                $3.16$ &         $3.57^{*}$ &                   $3.86$ &       $3.41$ \\
both songs         &        $3.49^{**}$ &   $3.29^{*}$ &  $3.47^{***}$ &          $2.78$ &                $3.28$ &         $3.61^{*}$ &               $3.93^{*}$ &  $3.61^{**}$ \\
\bottomrule
\end{tabular}
 \label{table:familiarity} 
 \end{table*}
 \setlength{\tabcolsep}{5pt}

\subsubsection{\textit{Interestingness} and quality metrics}
The $\mathit{interestingness}$ function is our computational means for assessing whether a segue foud by \dave{} is good or not. We would like to verify whether it agrees with the human perception of quality or not.
To this end, we compute the correlation of the \valence{} quality metrics and $\mathit{interestingness}$ for all of \dave{}'s segues that were used in the user trial.

We find that there is a statistically significant correlation between the \valence{} quality metrics and $\mathit{interestingness}$ for informative segues. This indicates that the $\mathit{interestingness}$ function, without being aware of semantics, relying only on statistical information and simple content descriptors, can successfully rate the quality of informative segues: on average, segues rated low by the trial participants are low also in $\mathit{interestingness}$, and vice versa. We believe that this is a very good result, given the complexity of the task. We observe worse results with funny segues, where we do not find any statistically significant correlations: deeper considerations might be needed, e.g.\ the role of semantics.
We also compute the correlation with \content{} quality metrics, but we do not find any strong statistically significant correlations. This is expected, since $\mathit{interestingness}$ is independent from the semantics of the segues. 

Finally, we turn to the correlation with \textm{} quality metrics. We do not find any correlation between $\mathit{interestingness}$ and \ww{}. This is as expected, since \ww{} does not depend directly on $\mathit{interestingness}$, but on $\mathit{path\_to\_text}$. But, we do find statistically significant correlation in informative segues for \understandable{} (0.22, $p$$<$$0.001$). This is an indication that $\mathit{interestingness}$, even though it is built around the concept of infrequency, does not favour obscure segues.

We report all the results in Table \ref{table:correlation}.

\begin{table}
\caption{\dave{}'s segues. Correlation of quality metrics and \textit{interestingness} score. $^{*}$: $p<0.05$; $^{**}$: $p<0.01$; $^{***}$: $p<0.001$.}
\begin{tabular}{lcc}
\toprule
{} & \multicolumn{2}{c}{\textit{interestingness}} \\ \midrule
{} & informative segues & funny segues \\
\midrule
\likeable{}       &       $\phantom{*}\phantom{*}0.25^{***}$ &       $0.13$ \\
\hq{}             &       $\phantom{*}\phantom{*}0.23^{***}$ &       $0.11$ \\
\si{}             &        $\phantom{*}\phantom{*}0.17^{**}\phantom{*}$ &      $0.00$ \\
\funny{}          &            $\phantom{*}\phantom{*}$$-0.06\phantom{*}\phantom{*}\phantom{*}\phantom{*}$ &  $-0.17^{*}$ \\
\informative{}    &         $\phantom{*}\phantom{*}0.14^{*}\phantom{*}\phantom{*}$ &       $0.04$ \\
\creative{}       &         $\phantom{*}\phantom{*}0.15^{*}\phantom{*}\phantom{*}$ &      $-0.09\phantom{*}$ \\
\understandable{} &       $\phantom{*}\phantom{*}0.22^{***}$ &       $0.10$ \\
\ww{}             &             $\phantom{*}\phantom{*}0.11\phantom{*}\phantom{*}\phantom{*}$ &       $0.05$ \\
\bottomrule
\end{tabular}
\label{table:correlation}
\end{table}

\section{Conclusions and Future Work} \label{sec:conclusions}
In this paper we introduced a novel method for generating item-to-item segues and implemented it in the case of song-to-song segues in a system called \dave{}. \dave{} can provide a wide variety of segues, that can be categorized as either funny or informative. The core of our method is $\mathit{interestingness}$, a domain-independent function for scoring the interestingness of paths in knowledge graphs. 

We evaluate \dave{} by means of a user trial, where we compare it against curated segues from a segment of the Radcliffe \& Maconie Show on BBC Radio 6 program, called \chain{}.
The use of \chain{} may be a limitation of this work. Segues from \chain{} have a peculiar style that fits the radio program, and are tailored to a particular kind of audience. They tend to be amusing and very creative. \dave{}, on the other hand, tends to be factual, and can only deliver funny segues made of word-play. In order to alleviate such problems, we compared funny segues and informative segues from the two methods separately. 
Notice too that, even though \chain{} has a high percentage of funny segues, this does not mean that its informative segues are weak: \chain{} is curated by experts and draw on the considerable knowledge of thousands of BBC listeners. 
In any case, the user trial was intended as a method to assess how \dave{}'s segues compare with curated segues from a 
trustworthy source --- segues that we can assume to be ``really good''. This gives a way of finding how far \dave{} is from being ``really good''. 
Our goal is not to demonstrate that our algorithm can be substituted for the listeners to the show, instead, we aim to provide an evaluation in a scenario where no algorithmic baseline is available.

We find that \dave{} can produce informative segues of the same quality, if not better, than \chain{}. We believe that this is an astonishing result, that gives an idea of the quality of our method. But, when turning to funny segues, the results are not as good. We believe that this is partially due to the \textit{interestingness} function, as we find evidence that it is much better suited to rate the quality of informative segues. Another reason might be that funny segues from \dave{} are limited to word-play, and this kind of humour does not appeal to everyone and may, in particular, not appeal to the participants in our user trial. In fact, even curated funny segues from \chain{} are not perceived as funny by participants in our trial. This may be a mismatch in sense of humour between listeners to the show and participants in the trial. It may also be due, in part, to the fact that segues are being read rather than being spoken. 
It is fair to say that, overall, the task of tackling the funny segues is only partly solved by the proposed model.

We find that good segues can spark interest in songs, and that this effect increases the more the user has high active engagement with music. Future directions might include the construction of a recommender system that incorporates segues to guide the discovery of new music, especially helping to solve the open problem of acceptance of recommendations for songs that deviate from user expectations (so-called divergent song recommendations \cite{Mehrotra2020}). 
We also notice that segues are particularly appreciated when they deal with music the user is already familiar with. Future work might therefore also concern the application of segues to enhance  recommendations for repeated consumption, a popular topic in the music domain \cite{Tsukuda2020}.

We are also interested in exploring the idea of learning the weights used by the  $interestingness$ scoring function, and in particular learning personalized weights that would adjust to fit the tastes of the user.

Other future directions involve the use of multiple segues to provide a narrative to accompany a playlist.

Lastly, we are interested in exploiting the domain-independent nature of our contribution, and to work with other kinds of items. For example, segues might be used by point-of-interest recommenders in the tourism domain to build a story around places to visit, going towards the idea of a virtual and intelligent tour guide.


\begin{acks}
We thank Peter Knees of TU Wien for useful discussions about the design of the user trial.

This publication has emanated from research conducted with the financial support of Science Foundation Ireland under Grant number 12/RC/2289-P2  which is co-funded under the European Regional Development Fund. For the purpose of Open Access, the authors have applied a CC BY public copyright licence to any Author Accepted Manuscript version arising from this submission.
\end{acks}

\bibliographystyle{ACM-Reference-Format}
\bibliography{sample-base}

\end{document}